\documentclass[%
 reprint,
superscriptaddress,
 amsmath,amssymb,
 aps,
 prl,
]{revtex4-2}

\usepackage{graphicx}
\setkeys{Gin}{keepaspectratio,width=1in,height=1in}
\usepackage{dcolumn}
\usepackage{bm}

\begin{document}


\title{Environmental Radiation Impact on Lifetimes and Quasiparticle Tunneling Rates of Fixed-Frequency Transmon Qubits}

\author{R.T. Gordon}
 \email{Corresponding author: Ryan.Gordon@ibm.com}
\author{C.E. Murray}
\author{C. Kurter}
\author{M. Sandberg}
\author{S.A. Hall}
\author{K. Balakrishnan}
\affiliation{IBM Quantum, T.J. Watson Research Center, Yorktown Heights, NY 10598}
\author{R. Shelby}
\affiliation{IBM Quantum, Almaden Research Center, San Jose, CA 95120}
\author{B. Wacaser}
\author{A. A. Stabile}
\author{J.W. Sleight}
\author{M. Brink}
\author{M.B. Rothwell}
\author{K. Rodbell}
\author{O.Dial}
\author{M. Steffen}
\affiliation{IBM Quantum, T.J. Watson Research Center, Yorktown Heights, NY 10598}





\date{\today}

\begin{abstract}
Quantum computing relies on the operation of qubits in an environment as free of noise as possible. This work reports on measuring the impact of environmental radiation on lifetimes of fixed frequency transmon qubits with various capacitor pad geometries by varying the amount of shielding used in the measurement space. It was found that the qubit lifetimes are robust against these shielding changes until the most extreme limit was tested without a mixing chamber shield in the refrigerator. In contrast, the quasiparticle tunneling rates were found to be extremely sensitive to all configurations tested, indicating these devices are not yet limited by losses related to superconducting quasiparticles.


\end{abstract}

\pacs{PACS numbers: }
\maketitle




Understanding the details of the interaction between environmental radiation and transmon qubits is critical for designing a measurement space that will support the development of quantum computing in superconducting qubits. Quasiparticle (QP) tunneling across Josephson junctions has been demonstrated to generate loss in qubits \cite{Martinis2009}, where the effect can be described as a change to the real part of the complex admittance \cite{Catelani2011a}.  QPs can be produced in superconducting metallization due to thermal and non-equilbirum sources, where the latter portion has been shown to dominate at the cryogenic temperatures associated with typical qubit operation \cite{Visser2011}.  These sources include infrared radiation \cite{Barends2011}, emission from radioactive impurities \cite{Vespalainen2020}, and cosmic rays \cite{McEwen2021}, which have been shown to create burst events within the underlying substrate \cite{Cardani2020, Wilen2020} transmitting phonons into the qubit environment \cite{Ioffe2004}.  Methods have been investigated to mitigate QP production, such as normal metal traps \cite{Riwar2016, Hosseinkhan2018}, modulation of the groundplane \cite{Grunhaupt2018, Henriques2019}, or superconducting 'gap engineering' of the leads in the vicinity of the junction \cite{Aumentado2004, Sun2012}.
Radiation at or near the qubit frequency, often near 5 GHz, would likely couple strongly to these devices and interfere with their operation, but even higher frequency radiation has also been proposed as problematic \cite{Serniak2019, Rafferty2021}. The superconducting gap of the metals used to fabricate qubits, near or above 100 GHz, may set another important energy scale, as this could entirely determine the population of non-equilibrium QPs that are generated in these systems. These QPs are predicted to cause degradation \cite{Catelani2011a} and determining how they currently limit the performance of transmon qubits is critical for building a complete picture of how coherence is impacted in these devices \cite{Catelani2011b, Wenner2013, Wang2014, Gustavsson2016}.

\begin{figure}
\includegraphics[height=0.375\textwidth,width=0.6\textwidth]{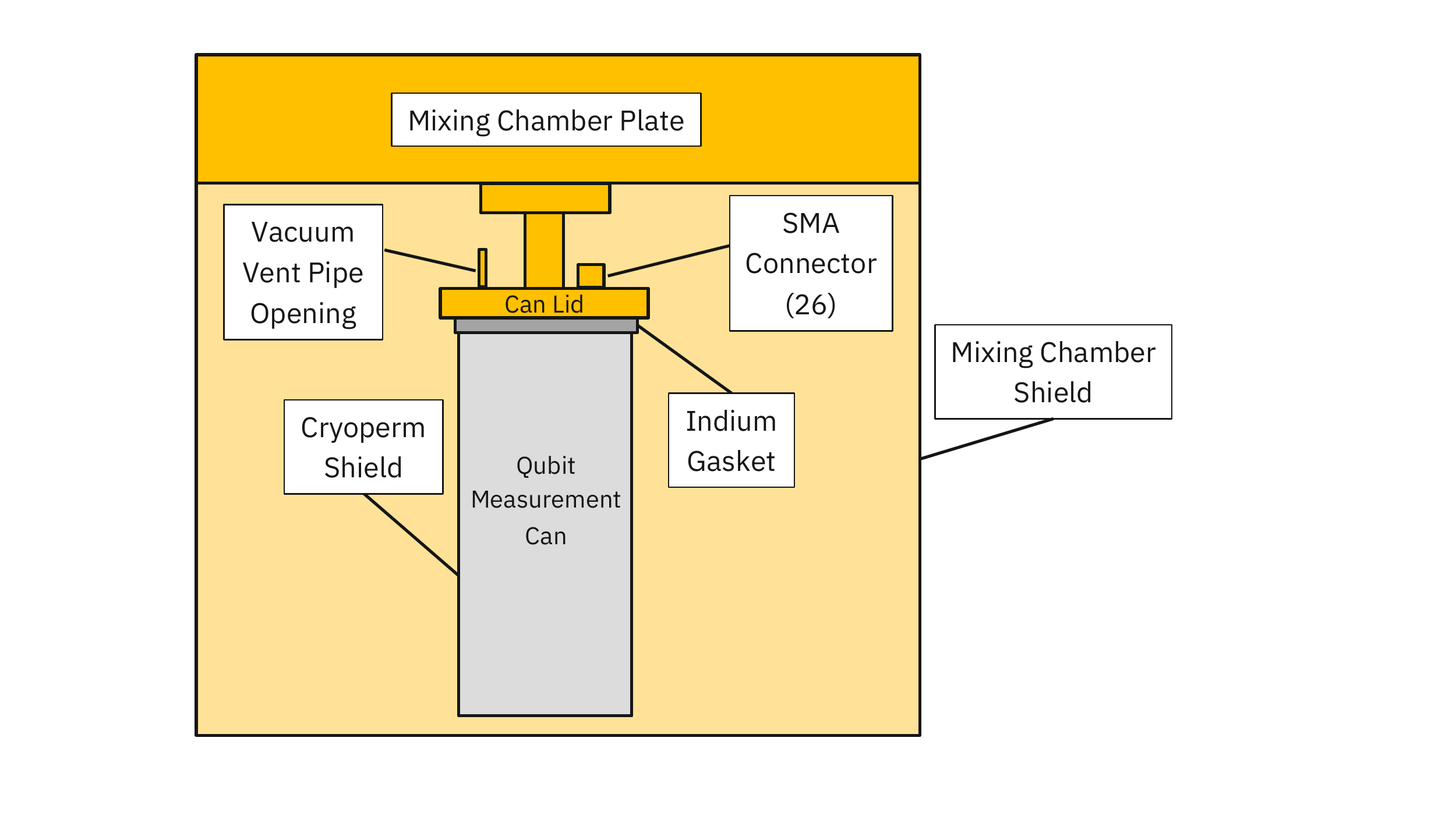}
\caption{\label{shields} Shielding elements used in the experiments.}
\end{figure}

This work focuses on learning how to properly shield the measurement space of fixed-frequency transmon qubits and to determine how these devices are presently limited by environmental radiation. In this study, the relaxation, decoherence, dephasing, and QP tunneling rates have been measured for a large ensemble of control devices. As illustrated in Fig.\ref{shields}, the shielding elements used in this study were: 1) an indium gasket at the measurement can lid to provide a light-tight environment for the devices, 2) a Cryoperm magnetic shield to provide shielding from both magnetic fields and electromagnetic radiation, 3) the mixing chamber shield of the dilution refrigerator itself used to isolate the mixing chamber from radiation originating at the higher temperature stages of the refrigerator, 4) a serpentine vacuum vent pipe installed through the can’s lid for removal of atmosphere from its interior, and finally 5) connector caps on the SMA inputs and outputs that pass through the lid of this measurement can. In the experiments performed, varying the measurement shielding has enabled us to control how much environmental radiation reaches the qubits, as well as for the case of Cryoperm, how much the magnetic field strength that reaches the devices is attenuated.

The main results of this work are that the QP tunneling rates of these qubits increased dramatically with the removal of each shielding element in the system but the relaxation, decoherence, and dephasing rates do not change much at all until the mixing chamber shield of the refrigerator is removed, corresponding to the most extreme environmental radiation limit used in this study. Additionally, in that high radiation limit, it was found that tapering the qubit capacitor paddles reduces the negative impact on performance from the environmental radiation.


Properties of the qubits used in these experiments are summarized in Table \ref{table1}, where the designs consist of rectangular shunting capacitor paddles \cite{Gambetta2017} fabricated on high-resistivity Si substrates with different coplanar gaps.  These gaps ranged from 1.5 microns to 70 microns, and the paddle metallization was refined to maintain a consistent value of capacitance and anharmonicity among the designs. 200 nm Nb films were sputter deposited, followed by lithographic patterning using a Nb etch.  Aluminum leads were then evaporated through e-beam lithography using a Dolan bridge approach \cite{Dolan1977} based on a poly(methyl methacrylate)/methyl methacrylate (PMMA/MMA) mask with two evaporation angles.  In between the two leads, the first aluminum layer received a 65 Torr-sec. exposure to oxygen to create the Josephson junction tunnel barrier.  In addition to the rectangular paddles, those with a tapered design in which the gap was reduced from the corners to the paddle centers to a distance of 6 microns were included to investigate the impact of the length of the Al leads.   Each Josephson junction was nominally the same for all five qubit designs but the capacitor pad geometry differed, as can be seen in the images of the qubits included in Table \ref{table1}. The size and shape of the capacitor pads for these devices may be critical for understanding how they interact with environmental radiation, so geometric dimensions for the pads are also included in this table for reference.

\begin{table}
\caption{\label{table1} Details of the transmon qubits used for this study, where qubit designs are referenced throughout using the notation here. Re[Y] refers to the real component of the simulated qubit admittance (see text).}
\begin{ruledtabular}
\begin{tabular}{cccccc}
\textrm{\vspace{0.001 in}}&
\textrm{\# 1}&
\textrm{\# 2}&
\textrm{\# 3}&
\textrm{\# 4}&
\textrm{\# 5}\\
\colrule
\vspace{0.001 in} & \includegraphics[height=0.1\textwidth,width=0.05\textwidth]{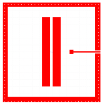} & \includegraphics[height=0.05\textwidth,width=0.05\textwidth]{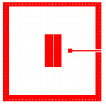} & \includegraphics[height=0.05\textwidth,width=0.05\textwidth]{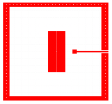} & \includegraphics[height=0.05\textwidth,width=0.05\textwidth]{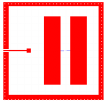} & \includegraphics[height=0.05\textwidth,width=0.05\textwidth]{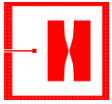}\\
Pad length ($\mu$m) & 500 & 240 & 300 & 500 & 480\\
Pad area (mm$^2$) & 0.06 & 0.029 & 0.036 & 0.12 & 0.114\\
Pad footprint (mm$^2$) & 0.07 & 0.03 & 0.03645 & 0.155 & 0.137\\
Gap size ($\mu$m) & 20 & 5 & 1.5 & 70 & 70 to 6\\
Re[Y]/Re[Y]$_{Design \#1}$ & 1 & 0.215 & 0.254 & 6.023 & 4.855\\
\end{tabular}
\end{ruledtabular}
\end{table}

Other features that differentiate these qubit designs are also included in Table \ref{table1}. Some additional key qualities relevant to distinguishing the qubit shunting capacitor pads are also included in Table \ref{table1}; such as the area of the capacitor pads, and the qubit footprint, which includes the area of the capacitor pads plus the area between them. 

All of the measurements for these experiments were performed in a BlueFors LD400 dilution refrigerator. Microwave characterization was done with a vector network analyzer and time domain qubit measurements were performed using both commercial and in-house sources. The input tones were sent into the fridge for readout and qubit control by passing through several cryogenic microwave attenuators at the various temperature stages of the refrigerator before reaching the devices. Connections to the device packaging were made using two 12-pin, single row Ardent connections. Output signals from the devices passed through a set of isolators, then through superconducting cables before reaching a HEMT amplifier located at the 3 K stage, and finally leaving the refrigerator to be amplified again at room-temperature, mixed down in frequency, and sent into a digitizer card for I-Q decomposition and readout analysis. 

Characterization of the qubit lifetimes was performed using conventional protocols. To determine the relaxation time, $T_1$, a $\pi$-pulse was used to prepare the qubit in the excited state, which was determined through Rabi oscillations. The qubit state was then read out after a variable delay time. This leads to an exponential decay of the qubit population vs. wait time, with the decay constant giving $T_1$. The decoherence time, $T_2$, was determined through Ramsey experiments, where from the excited state a $\pi$/2-pulse is applied placing the qubit in the $x-y$ plane of the Bloch sphere. After allowing the qubit state to precess a variable amount of time, a second $\pi$/2-pulse is applied to project the qubit state onto the $z$-axis. When the $\pi$/2-pulses are resonant with the qubit state, this results in a decaying envelope from which $T_2$ can be extracted. Dephasing times were calculated through the standard formulation $\frac{1}{T_2} = \frac{1}{2 T_1} + \frac{1}{T_\phi}$. QP tunneling measurements were conducted by analyzing the rate of charge parity shifts of the qubit state induced by a QP tunneling event \cite{Riste2013}, \cite{Serniak2018}. Specifically, a Ramsey-like pulse sequence was used to do this, where beginning from the excited state of the qubit, a $\pi$/2-pulse was used to project the state vector along the $y$-axis of the Bloch sphere. From there, the state vector was allowed to precess in the $x-y$ plane, where the direction of precession was dependent on the charge parity of the qubit, being either even or odd. After this, a second $\pi$/2-pulse was applied to project the qubit state onto the $z$-axis, allowing for the differentiation between even and odd charge parity. This technique was used to produce a map of the qubit's charge parity vs. time, from which a Fourier transform was used to create a power spectral density plot for these tunneling events. Finally, from the location of the "knee" in the power spectral density, one can extract a QP tunneling time, $T_P$, or rate, 1/$T_P$.

The packaging for the chips consisted of a printed circuit board (PCB) that was wire-bonded to twelve individual Si chiplets having approximately ten qubits each. The PCB was sandwiched between two thick gold-plated copper lids, where the chiplets sat in cavities milled out of one of these lids. Once assembled, this package was firmly mounted from a thick, gold-plated copper finger that was suspended from the mixing chamber of the dilution refrigerator. Surrounding this package and cold finger from below was a Cryoperm shield shaped like a can with an open top and closed bottom that mated with a lid to seal off the measurement space (see Figure \ref{shields}). The seal between the Cryoperm shield and the lid was filled with a pre-formed indium gasket in the shape of a ring having a thickness of 1.5 mm, outer diameter of 85 mm, and inner diameter of 73.2 mm. The intended purpose of the gasket was to form a light-tight seal at the Cryoperm-lid interface. Surrounding the entire experiment at the mixing chamber of the refrigerator was a gold-plated copper mixing chamber shield to keep out radiation from the upper temperature stages.

\begin{figure}
\includegraphics[height=0.3\textwidth,width=0.6\textwidth]{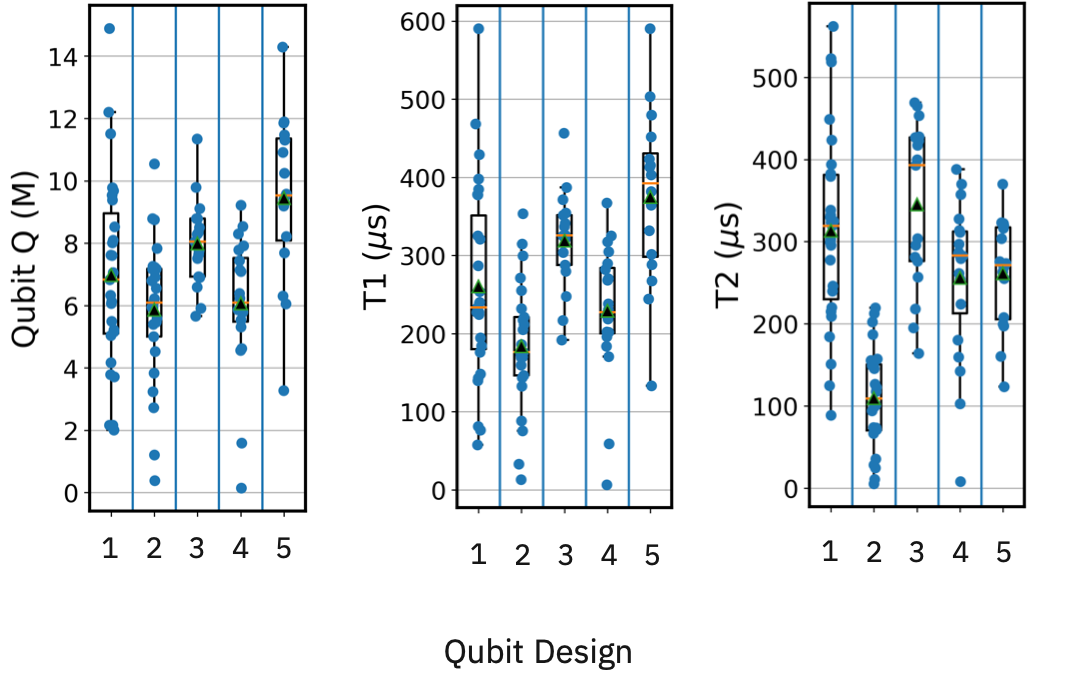}
\caption{\label{qubit_char} Qubit quality factor, $T_1$, and $T_2$ for the transmon qubits used in this study.}
\end{figure}


To understand best the impact of environmental radiation on qubit performance, select devices were chosen to form the ensemble for these experiments. See Table \ref{table1} for a summary of physical parameters for the five different qubit designs contained in this ensemble. The coherence measurements for these devices are reported in Figure \ref{qubit_char}. The individual data values for each device reported in this manuscript for rates and quality factors represent averages from approximately 30 measurements taken over an eight-hour period. The T$_1$ data for the best performing design (\#5) had a median value of nearly 400 $\mu$s, and highest value at nearly 600 $\mu$s, with corresponding qubit Q of over 14 million. For T$_2$, the median was around 320 $\mu$s for the best performing device design (\#1), with highest values near 550 $\mu$s. A summary of the qubit quality factor, $T_1$, and $T_2$ values for the transmon qubits used in these experiments is shown in Figure \ref{qubit_char}.

\begin{figure}
\includegraphics[height=0.3\textwidth,width=0.6\textwidth]{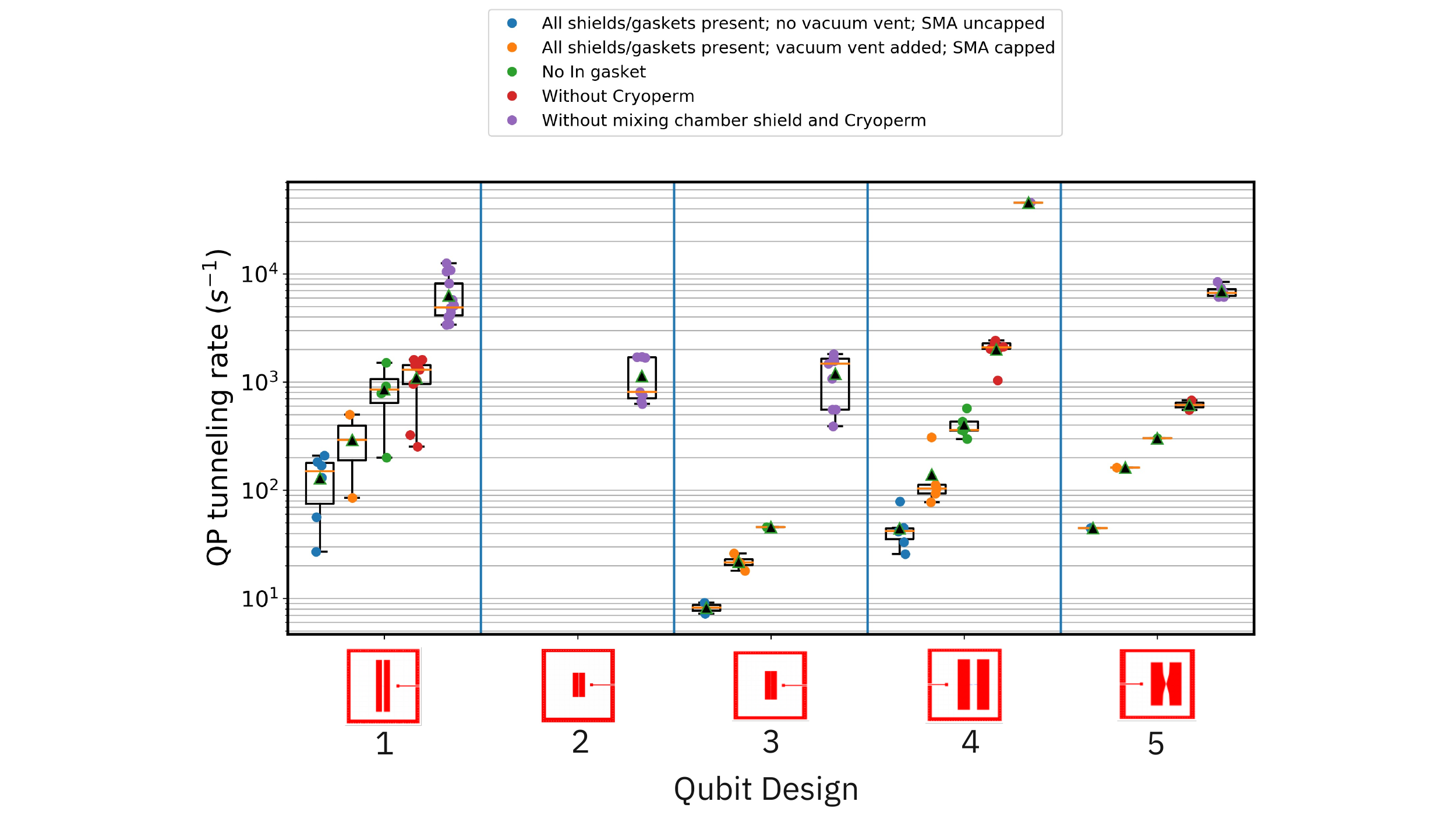}
\caption{\label{QP_various_shields} QP tunneling rates for the five different qubit designs measured in five different shielding conditions. Limited statistics for design \#2 are due to difficulty in measuring from charge noise fluctuations and readout fidelity improvements.}
\end{figure}

The environmental radiation level in the qubit measurement space for these experiments was varied systematically by removing each shielding element from the measurement configuration one at a time. It should be noted that removal of the Cryoperm also increases the magnetic field intensity in the device measurement space for this setup by a factor of approximately 1000 at room-temperature, determined by Hall probe measurements \cite{Rode1983}, in addition to also likely increasing radiation levels when removed. A drawing of the shielding elements varied in this study and how they are physically positioned during the measurements is shown in Figure \ref{shields}. The shielding elements were: 1) an indium gasket at the seal between the Cryoperm can and its lid, 2) a Cryoperm shield, 3) the mixing chamber shield of the dilution refrigerator, 4) a vacuum vent pipe passing through the can lid, and 5) caps on the two unused SMA connectors at the can lid that were generally for microwave inputs and outputs.

\begin{figure}
\includegraphics[height=0.5\textwidth,width=0.5\textwidth]{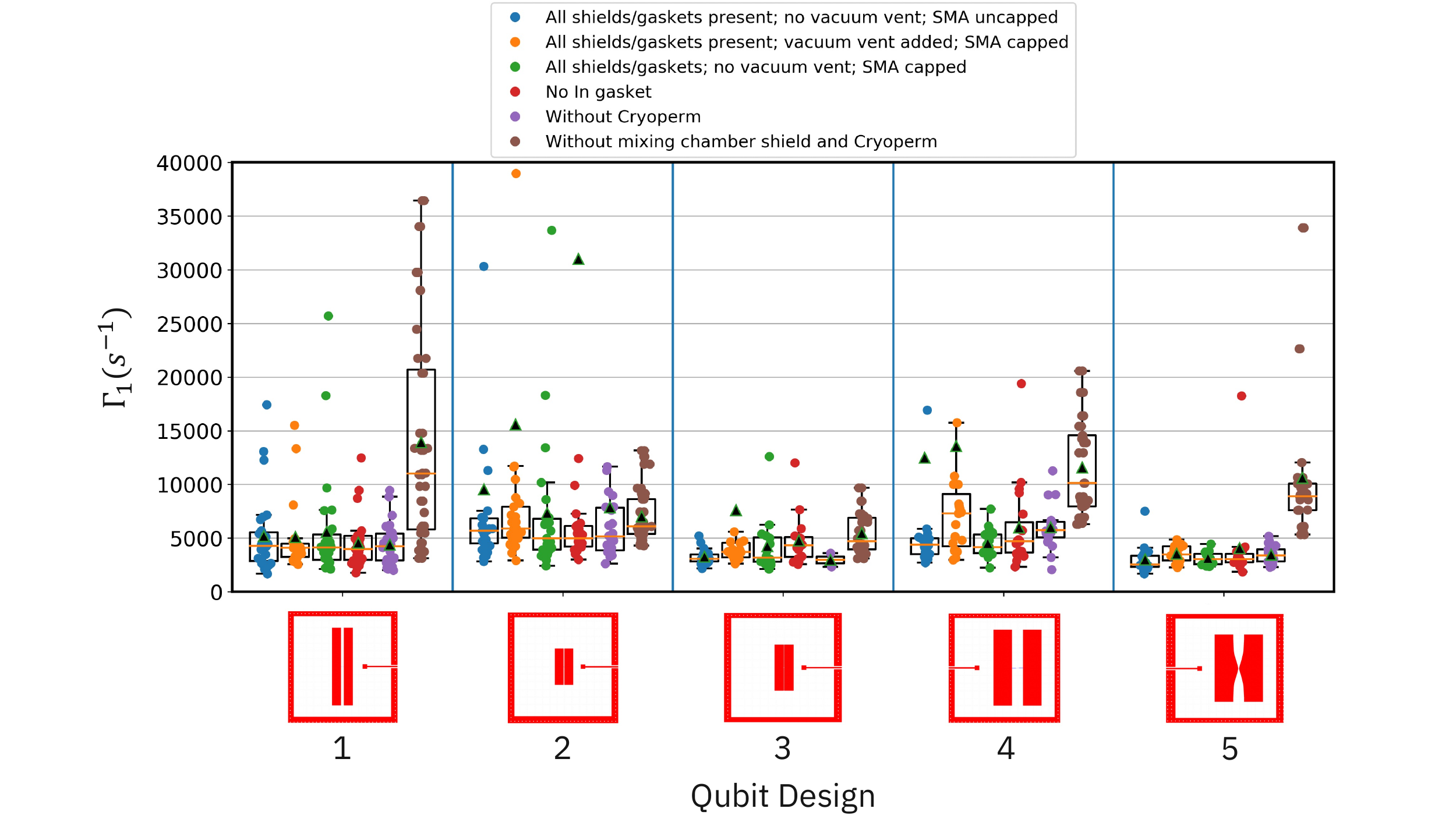}
\caption{\label{relax_rate} Relaxation rate, $\Gamma_1=1/T_1$ , for the six different shielding configurations studied with the same set of transmon qubits.}
\end{figure}

\begin{figure}
\includegraphics[height=0.5\textwidth,width=0.5\textwidth]{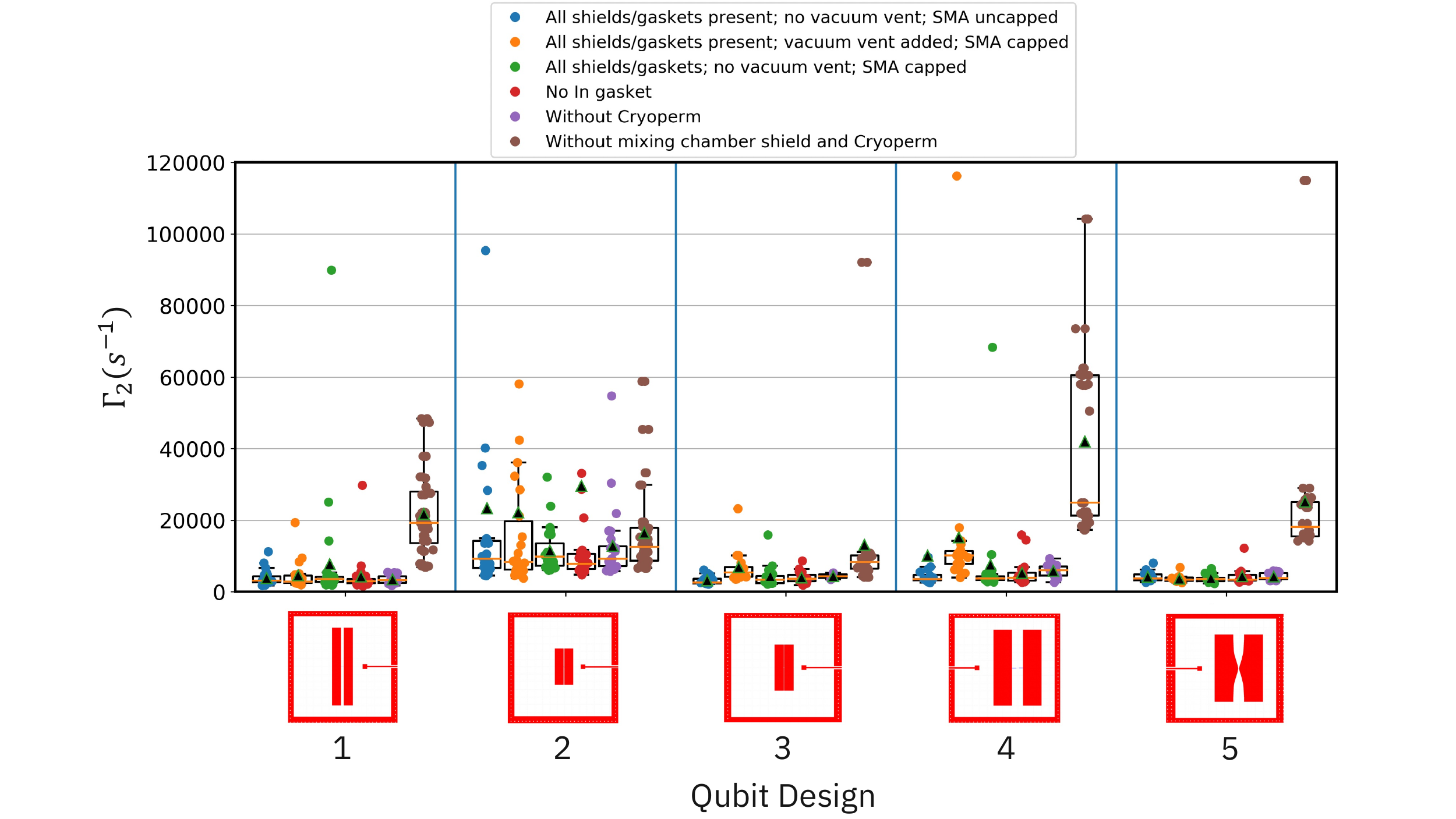}
\caption{\label{decoh_rate} Decoherence rate, $\Gamma_2=1/T_2$ , for the six measurement configurations studied with the same set of transmon qubits.}
\end{figure}

\begin{figure}
\includegraphics[height=0.5\textwidth,width=0.5\textwidth]{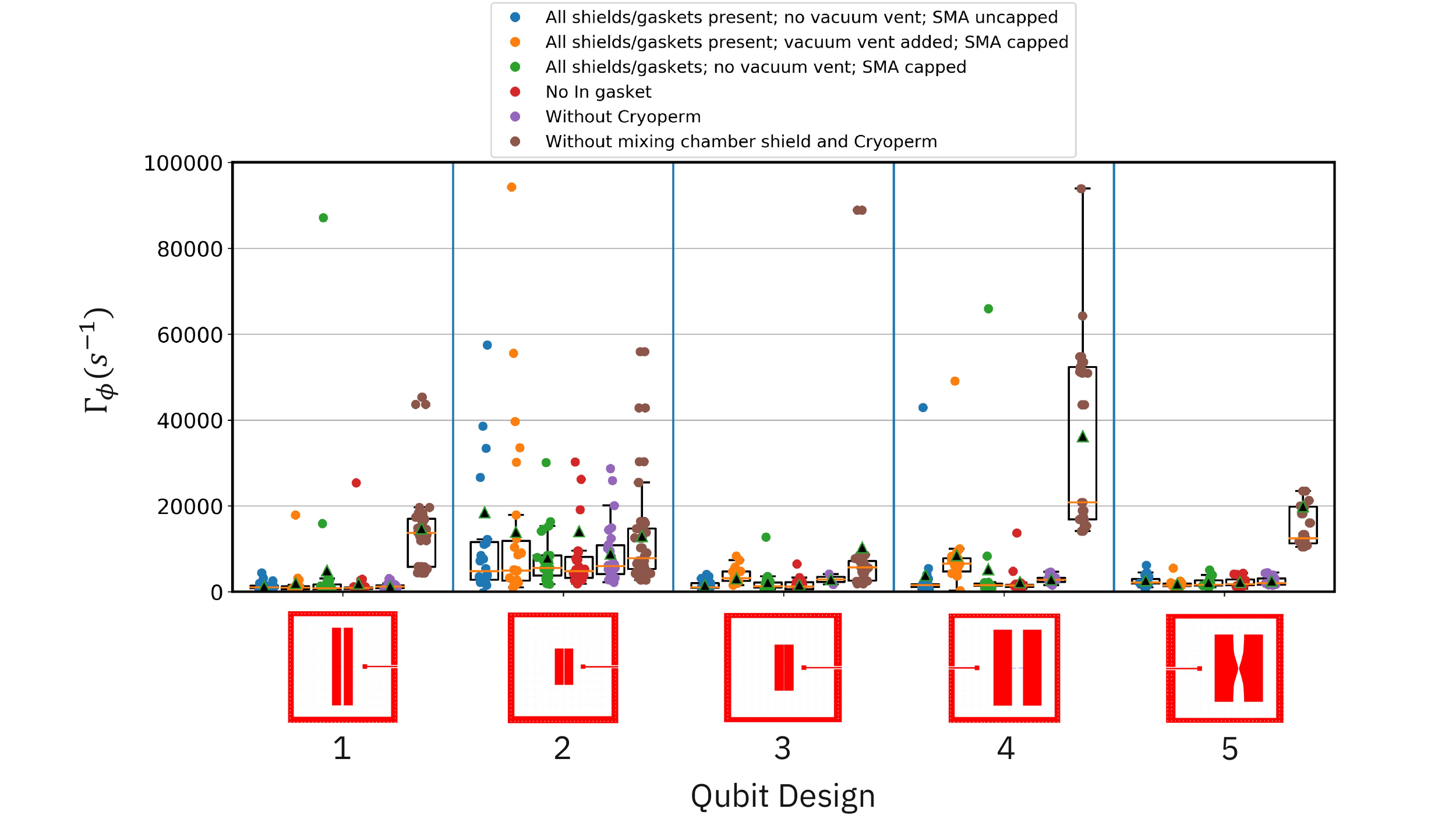}
\caption{\label{dephasing_rate} Dephasing rate, $\Gamma_{\phi}=1/T_{\phi}$, for the six configurations studied.}
\end{figure}

Removing the indium gasket at the can-lid interface breaks the light-tight seal that was surrounding the devices, allowing environmental radiation to leak into the qubit measurement space. The impact of removing this gasket is most easily seen on the QP tunneling rates collected throughout these experiments, where in Figure \ref{QP_various_shields} one can plainly see the increase associated with the removal of the gasket (green data). This rate increase observed after removing the gasket is likely due to pair-breaking effects associated with this environmental radiation. To quantify the observed increase, for qubit design \#1, the median  tunneling rate increased from 181.2 $s^{-1}$ to 850.4 $s^{-1}$, as can be seen in Table \ref{table2}. In contrast, there is very little impact from removing the gasket on the rates for relaxation (Figure \ref{relax_rate}), decoherence (Figure ~\ref{decoh_rate}), and dephasing (Figure \ref{dephasing_rate}), indicating that environmental radiation has a much stronger influence over QP tunneling than it has on qubit lifetimes \cite{Serniak2019}.

\begin{figure}
\includegraphics[height=0.5\textwidth,width=0.5\textwidth]{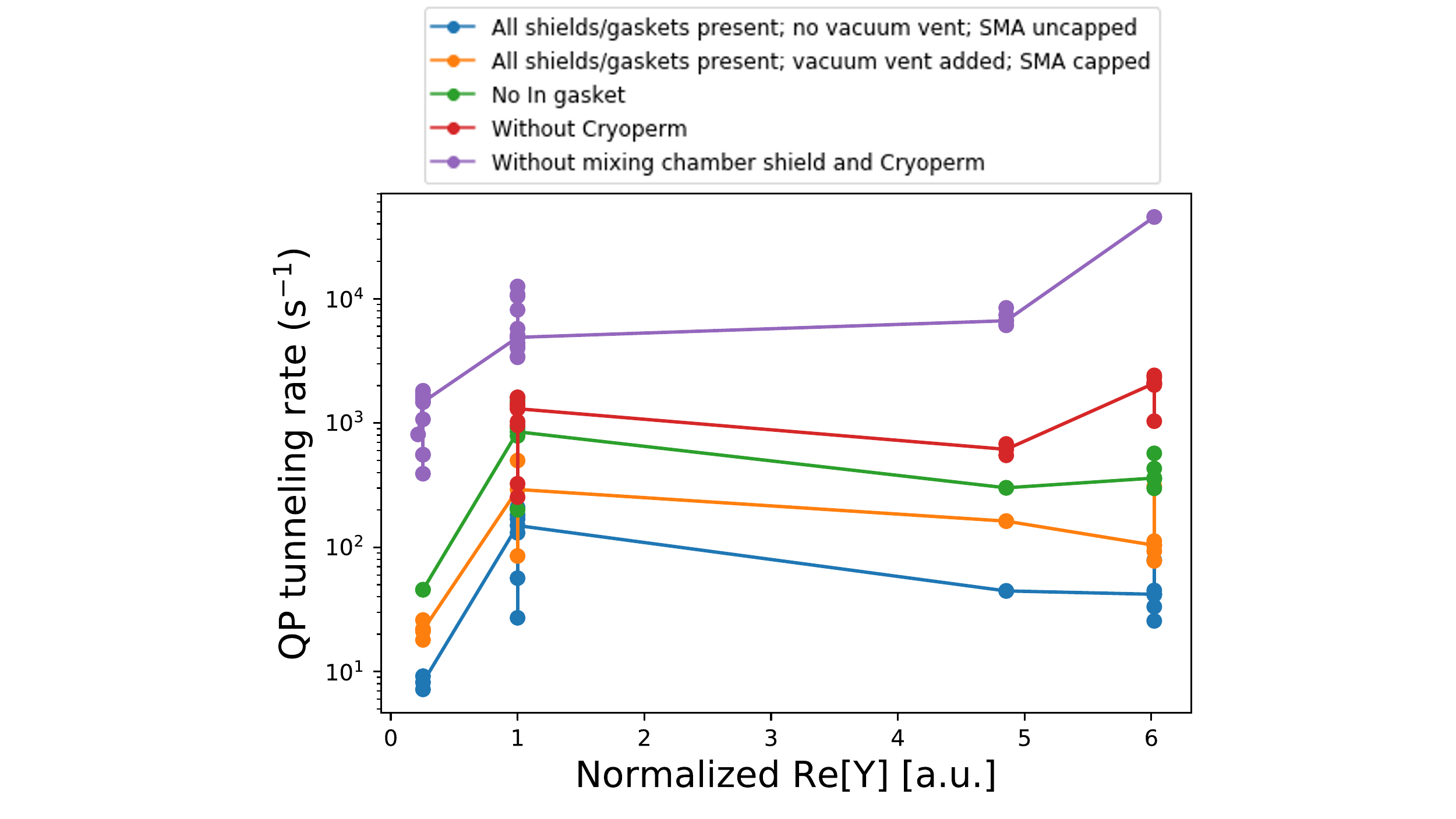}
\caption{\label{QP_rate_vs_pad} QP tunneling rate vs. normalized admittance in arbitrary units, with values normalized by the admittance calculated for Design \#1. In the legend, SMA refers to the type of connector at the qubit can lid that was measured both capped and uncapped.}
\end{figure}

The Cryoperm shield was also removed systematically from around the qubits. As for its effect on the qubit lifetimes the results are shown by visualizing the relaxation rates (Figure \ref{relax_rate}), decoherence rates (Figure \ref{decoh_rate}), and dephasing rates (Figure \ref{dephasing_rate}). From this data (purple symbols), one can see that there is little impact on any of these three rates due to the removal of the Cryoperm shield. However, an examination of the corresponding QP tunneling rates shows a different result, illustrated in Figure \ref{QP_various_shields} and Figure \ref{QP_rate_vs_pad}. The Cryoperm shield in principle can serve two purposes to improve the qubit measurements: firstly, to serve as another radiation shield; and secondly, to attenuate magnetic fields inside the can. It may not be too surprising that the rates in Figure \ref{relax_rate}, Figure \ref{decoh_rate}, and Figure \ref{dephasing_rate} show little impact, if any, from the removal of the Cryoperm since removing the light-tight gasket had little effect. What may be less obvious is the lack of degradation due to Abrikosov vortices in the superconducting films of the devices \cite{Abrikosov}, since the magnetic field at the sample space likely increased dramatically due to the Cryoperm removal.

Perhaps the most interesting change in the shielding configuration used for these measurements is illustrated by the brown data shown in Figure \ref{relax_rate}, Figure \ref{decoh_rate}, and Figure \ref{dephasing_rate}; corresponding to removal of the mixing chamber shield of the dilution refrigerator, while also not including the Cryoperm can. For some qubit designs, this change has led to a dramatic increase in the measured relaxation, decoherence, and dephasing rates. In particular, the designs that show the highest rate increases, and therefore the most degradation in performance, are those with the largest capacitor pad area, with exact values listed in Table \ref{table1} for all five designs. Specifically, designs \# 1, \# 4, and \# 5 show the largest increase in median rates. Another interesting observation is that the qubits employing tapered paddles (\#5) were impacted much less than those with standard paddle geometries (\#4), for pad areas that are quite similar. The tapered design may somehow reduce the associated loss from this effect by changing the characteristics of the current that flows through the pad to the Josephson junction of the device.

\begin{table}
\caption{\label{table2} Median rates for QP tunneling, relaxation, decoherence, and dephasing of qubit design \#1 for various shielding configurations.}
\begin{ruledtabular}
\begin{tabular}{ccccc}
\textrm{\textbf{Shielding}}&
\textrm{$\Gamma_{QP}$($s^{-1}$)}&
\textrm{$\Gamma_{1}$($s^{-1}$)}&
\textrm{$\Gamma_{2}$($s^{-1}$)}&
\textrm{$\Gamma_{\phi}$($s^{-1}$)}\\
\colrule
All Shields; capped SMA & 181.2 & 4118.2 & 3538.3 & 845.4\\
Vacuum pipe added & 292.2 & - & - & -\\
No In lid gasket & 850.4 & 3972.2 & 2931.5 & 865.3\\
No Cryoperm shield & 1303.2 & 4253.6 & 3286.7 & 1177.7\\
No MXC/Cryoperm shield & 4743.0 & 11001.8 & 19266.0 & 13668.7\\
\end{tabular}
\end{ruledtabular}
\end{table}

The QP tunneling rates for this extreme radiation case increased by as much as two orders of magnitude after removal of the mixing chamber shield, while also having no Cryoperm shielding, shown by the purple data in Figure \ref{QP_various_shields}. The qubit designs with the largest areas in Figure \ref{QP_various_shields} (design \#s 1, 4, and 5) also clearly show the highest QP tunneling rate increases upon removal of the mixing chamber shield. This observation is consistent with a simple model where the qubit is considered as an antenna receiver for environmental radiation \cite{Rafferty2021}. Larger qubit pad areas absorb more radiation from the envrionment than do small ones and this radiation is likely pair-breaking for superconductivity, either through a direct mechanism or through an indirect one. Another feature of this data to notice is that the median QP tunneling rates for the largest qubit designs shown in Figure \ref{QP_various_shields} are very near 10,000 $s^{-1}$, as are the median relaxation rates for the same qubit designs shown in Figure \ref{relax_rate}, perhaps indicating that in this high radiation regime the qubit relaxation rates are limited by QP-related  losses.

In order to assess the susceptibility of different qubit designs to incident radiation, EM simulations were conducted using HFSS (Canonsburg, PA).  Analogous to Lorentz reciprocity, the properties of the qubit paddles as an antenna \cite{Rafferty2021} radiating into a lossy environment were utilized to establish the qubit admittance, Y, where its real component, Re[Y],
is proportional to the relaxation rate $\Gamma_1 = \frac{Re[Y]}{C_q}$ \cite{Houck2008} and $C_q$ represents the total capacitance of the qubit (see Supplementary Information).
QP tunneling rates as a function of Re[Y], normalized by the value corresponding to Design \#1, are displayed in Figure \ref{QP_rate_vs_pad}.  One observes a clear correlation that qubit paddles with larger footprints exhibit greater QP tunneling events.  The dependence between the qubit footprint and Re[Y] is nonlinear suggesting that it is not simply the area of the paddle metallization that dictates the susceptibility of a given design to QP events.  It is of interest that the tapered qubit designs possess a much lower QP tunneling rate than their corresponding non-tapered paddles for a given paddle gap.  These results are consistent with those of \cite{Wang2014}, who demonstrated that QP lifetimes are greater within transmon qubit designs that possess long leads, in contrast to devices with large paddle regions that can trap the QPs.  Those measurements were conducted on 3D Al qubits on sapphire substrates, suggesting that trapping mechanisms, such as vortices, were responsible \cite{Wang2014}.

Minor shielding elements were also varied in this study. One of these was the vacuum vent tube that passed through the lid of the measurement can. Design \# 4 did show a slight degradation of qubit lifetimes from the addition of the vacuum vent, seen in Figure \ref{relax_rate}, Figure \ref{decoh_rate}, and Figure \ref{dephasing_rate}, indicating that some feature of this design may couple to environmental radiation leaking in through this pipe. The QP rates increased for all designs studied, shown in Figure \ref{QP_rate_vs_pad}. The second minor shielding element varied was the addition of cover caps to unused SMA connectors for inputs and outputs that pass through the measurement can’s lid. A small change in the QP rate was observed (Figure \ref{QP_rate_vs_pad}) but there was little impact on the relaxation rates (Figure \ref{relax_rate}), decoherence rates (Figure \ref{decoh_rate}), or dephasing rates (Figure \ref{dephasing_rate}) for these qubits.


State-of-the-art, fixed-frequency, transmon qubits having five different shunting capacitor pad designs and median Q values ranging from 6.1 to 9.5 million were used as control devices to study the impact of environmental radiation on both qubit lifetimes and QP tunneling rates. The environmental radiation was varied in a controlled way by changing the shielding configuration used around the measurement space. The shielding elements that were varied throughout this study were: 1) an indium gasket seal at the qubit measurement can lid, 2) the Cryoperm shield of the measurement can, 3) the mixing chamber shield of the dilution refrigerator, 4) a vacuum vent pipe at the can’s lid, and 5) connector caps to cover the SMA input and output connectors through this lid.

It was found that the relaxation, decoherence, and dephasing rates of these transmon qubits were surprisingly robust to changes in the environmental radiation and only when the mixing chamber shield of the refrigerator was removed, along with the Cryoperm shield, did these rates show a significant increase. For this most extreme radiation limit, the devices with the largest capacitor pad areas were most susceptible to the environmental radiation (Design \#s 1, 4, and 5). When comparing the tapered and standard device designs, which have very nearly the same pad area, it was discovered that the tapered design has less degradation from this radiation.

In contrast, the QP tunneling rates of these qubits changed dramatically with these shielding configuration changes. Again, the devices with the largest capacitor pad areas had the most sensitivity to environmental radiation, indicating that the radiation absorbed by the qubit is pair-breaking for superconductivity.

Finally, perhaps the most significant result from these experiments is that by changing the environmental radiation through these shielding configurations, the QP tunneling rates were able to be increased in a controlled manner without affecting the relaxation, decoherence, and dephasing rates of the qubits until the most extreme radiation limit was reached, where the devices were subjected to relatively high levels of blackbody radiation due to removal of the mixing chamber shield of the refrigerator. This result is critically important because it shows that the highest coherence transmon qubits are not yet limited by QP  tunneling processes when properly shielded in their measurement environment.

\begin{acknowledgments}
The authors would like to thank Robert Sandstrom and Kevin Stawiasz for help with many aspects of the experiment, Khan Dang for wire-bonding work on the packages, Serafino (Sonny) Carri for help with preparing the low-temperature environment, and Kenny Tran for various technical assistance.  Samples were fabricated using the IBM Microelectronics Laboratory and the IBM Central Scientific Services facility.
\end{acknowledgments}

\section{Supplementary Information}

The qubit admittance was calculated using Ansys HFSS (Canonsburg, PA) by generating qubit paddles on a dielectric substrate which is surrounded by a radiation box possessing a finite conductivity.  As mentioned in the manuscript text, we utilize the reciprocity between the interaction of a lossy enclosure with the radiation emanating outward from a qubit and a qubit paddle configuration as a receiving antenna to incident radiation to quantify the admittance.




\end{document}